\documentclass[sigconf,authorversion]{acmart}

\usepackage{booktabs}
\usepackage{multirow}
\usepackage[official]{eurosym}
\usepackage{listings}
\usepackage{enumitem}
\usepackage{balance}

\begin{document}

\copyrightyear{2021} 
\acmYear{2021} 
\setcopyright{acmcopyright}\acmConference[SIGIR '21]{Proceedings of the 44th International ACM SIGIR Conference on Research and Development in Information Retrieval}{July 11--15, 2021}{Virtual Event, Canada}
\acmBooktitle{Proceedings of the 44th International ACM SIGIR Conference on Research and Development in Information Retrieval (SIGIR '21), July 11--15, 2021, Virtual Event, Canada}
\acmPrice{15.00}
\acmDOI{10.1145/3404835.3463243}
\acmISBN{978-1-4503-8037-9/21/07}

\fancyhead{}
\title{POINT\-REC: A Test Collection for Narrative-driven Point of Interest Recommendation}

\author{Jafar Afzali}
\affiliation{%
  \institution{University of Stavanger}
}
\email{j.afzali@stud.uis.no}

\author{Aleksander Mark Drzewiecki}
\affiliation{%
  \institution{University of Stavanger}
}
\email{am.drzewiecki@stud.uis.no}

\author{Krisztian Balog}
\affiliation{%
  \institution{University of Stavanger}
}
\email{krisztian.balog@uis.no}

\begin{abstract}
This paper presents a test collection for contextual point of interest (POI) recommendation in a narrative-driven scenario.  There, user history is not available, instead, user requests are described in natural language.
The requests in our collection are manually collected from social sharing websites, and are annotated with various types of metadata, including location, categories, constraints, and example POIs.
These requests are to be resolved from a dataset of POIs, which are collected from a popular online directory, and are further linked to a geographical knowledge base and enriched with relevant web snippets.
Graded relevance assessments are collected using crowdsourcing, by pooling both manual and automatic recommendations, where the latter serve as baselines for future performance comparison. 
This resource supports the development of novel approaches for end-to-end POI recommendation as well as for specific semantic annotation tasks on natural language requests.
\end{abstract}



\maketitle

\section{Introduction}

Point of interest (POI) recommendation entails the task of recommending locations (e.g., restaurants, arts, and entertainment) to users based on their interest profiles.
It is a particular item recommendation task, which has attracted considerable interest over the past decade.
POI recommendation has especially gained popularity with the emergence of location-based social networks (LBSNs), like Yelp, Gowalla, or Foursquare, and is seen as an essential feature of those platforms~\citep{Ye:2011:EGI,Cheng:2012:FMF,Cheng:2013:WYL}.
On LBSNs, users share their social experience via check-ins and ratings/reviews.  This data \emph{implicitly} captures their preferences on locations and provides the basis for future recommendations.  Crucially, these general preferences are often insufficient when making recommendations; context may introduce additional constraints, not necessarily inclined to personal preferences~\citep{Aliannejadi:2018:PCP}.  For example, a person may be a frequent visitor to nightlife spots, but would prefer very different type of places during a weekend family getaway.

Aiding POI recommendation by incorporating contextual information was the very focus of the Contextual Suggestions track at the Text Retrieval Conference (TREC)~\citep{DeanHall:2015:OTC,Hashemi:2016:OTC}.
There, systems are required to make suggestions for a particular person in a specific context (e.g., a city and trip type) based on user profiles, which comprise stated preferences on POIs in different cities and contexts.
The particular needs and preferences of users are supposed to be \emph{inferred} from their interests.
This, however, appears to be a challenge, and ``not many successful participants took into account context in their proposed approaches''~\citep{Aliannejadi:2018:PCP}.
Our objective with this work is to eliminate difficulties associated with the extraction of contextual preferences and constraints, by making these more explicit, thereby facilitating research on novel contextual POI recommendation methods. 

This paper addresses a POI recommendation scenario, where a contextual information need is stated \emph{explicitly} in natural language.  This scenario has been termed as \emph{narrative-driven recommendation} in \citep{Bogers:2017:DSN}.  
However, unlike in~\citep{Bogers:2017:DSN}, past user transactions are not available in our setting (nor is user profile information, apart from what is explicitly stated in the description).  This renders it a particularly difficult and interesting problem.  It may be argued that this is more of a \emph{search} task than a recommendation one, since there is an explicit expression of the user need as opposed to implicit preferences captured by user history.  At the same time, the user is asking for \emph{recommendations}, i.e., they are seeking for suggestions to choose from.
Their information needs are too complex to be expressed as search queries that would be understood by search engines.  Rather, these are exactly the types of requests that one would make to a conversational assistant.
One particularly challenging aspect of this problem is natural language understanding, i.e., extracting and interpreting preferences that are expressed in free text form.  
Our work facilitates research in this area by providing semantically rich ground truth annotations of user preferences. 

We present the POINTREC collection consisting of the following:
\begin{itemize}[leftmargin=2em]
	\item A collection of 695K POIs, spanning over 23 countries and over 2K cities, which are collected from a popular online directory (Yelp), and are further linked to a geographical knowledge base (GeoNames) and enriched with relevant web snippets.
	\item A set of 112 information needs originating from real users, each with a text description detailing the context and preferences of the user.  These natural language descriptions are enriched with metadata annotations and, if possible, linked to structured knowledge repositories (including category, group, location information, constraints, and examples).
	\item Graded relevance assessments collected via crowdsourcing.
\end{itemize}

\noindent
It is worth pointing out that collaborative filtering techniques, which have been the predominant approach to POI recommendation to date~\citep{Ye:2011:EGI,Cheng:2012:FMF,Liu:2013:PPR,Lian:2014:GJG,Lian:2018:GSL}, are not applicable in our scenario, given the lack of user-item interactions.
Instead, we approach the task using a retrieval-based approach, where constraints are expressed as structured queries on fielded representations of items.
We offer a starting point by implementing and comparing the performance of three variants of this baseline.

The POINTREC collection and our baselines are made available at \url{https://github.com/iai-group/sigir2021-pointrec}.

\vspace*{-0.5\baselineskip}
\section{Related work}

The emergence of location-based social networks (LBSNs) brought about the need for POI recommendation as an essential service, in order to help users discover and explore new places~\citep{Ye:2011:EGI,Cheng:2012:FMF,Cheng:2013:WYL}.
POI recommendation in LBSNs ``distinguishes itself from traditional item recommendation [...] via geographical influence among POIs''~\citep{Wang:2018:EPG}.
While traditional collaborative filtering approaches offer a natural starting point, these need to be extended to incorporate spatial, temporal, and social information~\citep{Ye:2011:EGI,Gao:2013:ETE,Yuan:2013:TPR,Yang:2019:RUM}.
Another major challenge is data sparsity.  While there are millions of POIs, each user visits only a handful of them, making the user-POI interaction matrix extremely sparse~\citep{Yin:2016:JMU}.  Latent factor models, and in particular matrix factorization approaches, are often adopted to alleviate data sparsity~\citep{Cheng:2012:FMF,Liu:2013:PPR,Lian:2014:GJG,Lian:2018:GSL,Rahmani:2020:JGT}.  Alternatively, probabilistic generative models may also be employed~\citep{Yin:2016:JMU}.  Most recent work has shown the effectiveness of embedding-based approaches for POI recommendation~\citep{Rahmani:2019:CLE,Yang:2019:RUM,Yang:2020:LHH}.
Evaluation datasets are created by crawling LBSN data  via public APIs; these crawls are often shared by individual researchers, e.g., for Gowalla~\citep{Cho:2011:FMU,Cheng:2012:FMF,Liu:2017:EEP}
and Foursquare~\citep{Yuan:2013:TPR,Gao:2015:CPI,Yang:2016:PCM}.
These datasets are often restricted to specific geographic regions.

A key characteristic that makes POI recommendation a fertile ground for research is the inherent need for contextualization.  Context may be a previous POI (or category of POI)~\citep{Liu:2013:PPR,Benetka:2017:AIN}, temporal state (e.g., work vs. leisure time)~\citep{Yuan:2013:TPR,Rahmani:2020:JGT}, travel time~\citep{Braunhofer:2015:CMP}, or weather~\citep{Braunhofer:2015:CMP}.  Recommendations may even be shown proactively to users when the inferred context seems appropriate~\citep{Braunhofer:2015:CMP,Benetka:2017:AIN}.
The contextual appropriateness of locations may be estimated using heuristics~\citep{Braunhofer:2015:CMP} or cast as a binary classification problem~\citep{Aliannejadi:2018:PCP}.

The TREC Contextual Suggestions (TREC CS) track~\citep{DeanHall:2015:OTC,Hashemi:2016:OTC} provides a reusable test collection by considering five contextual dimensions: city (e.g., Boston), trip type (e.g., holiday), duration (e.g., weekend trip), type of group (e.g., friends), and a season the trip will occur in (e.g., summer).  The POI collection covers attractions from 272 North American cities.  POIs are represented by their URL, title, and city ID.
There are several key differences between TREC-CS and our POINTREC test collection:
\begin{itemize}[leftmargin=2em]
	\item The information needs at TREC CS are ``synthetic'' in the sense that they are generated by crowdsourcing using an imaginary scenario.  Ours, on the other hand, are real information needs by real people with an actual intent to travel.
	\item The contexts of information needs as well as user preferences in our collection are described in natural language in detail, and have been manually annotated along several dimensions; in case of TREC CS, user preferences need to be inferred.  On the other hand, our collection does not include past user history (unless references are made to prior experiences in the description), thereby making it a cold start problem.
	\item Our POI collection covers 23 countries, while TREC CS deals with North American cities only.
\end{itemize}

\vspace*{-0.25\baselineskip}
\section{Information needs}
\label{sec:infoneeds}

POINTREC comprises 112 information needs, spanning over four main categories (Section~\ref{sec:infoneeds:categories}).  These represent actual information needs of different people, which have been collected from community question answering and forum sites (Section~\ref{sec:infoneeds:collect}).  Each of these requests has been annotated with a rich set of metadata attributes (Section~\ref{sec:infoneeds:metadata}).  Both these steps have been performed manually by the authors of the paper to ensure high data quality.  The resulting collection exhibits a diverse set of needs and attributes (Section~\ref{sec:infoneeds:stat}).

\vspace*{-0.25\baselineskip}
\subsection{Categories}
\label{sec:infoneeds:categories}
\vspace*{-0.25\baselineskip}

We collected information needs in four main categories: \textit{Active life}, \textit{Arts and entertainment}, \textit{Restaurants and food}, and \textit{Nightlife}.  This particular selection was motivated by looking at the most popular top-level categories in Yelp.\footnote{Actually, \textit{Restaurants} and \textit{Food} are two separate categories in Yelp, which have been merged into a single one here.}

Within these main categories, a number of sub-categories are distinguished, following Yelp's category system.\footnote{\url{https://www.yelp.com/developers/documentation/v3/all_category_list}} Due to inappropriateness, a few sub-categories targeting adult entertainment have been actively avoided. 
A sample from the list of subcategories is given in Table~\ref{tab:categories}.

\begin{table*}[t]
	\caption{Category system, with a sample of subcategories for each main category.}
	\vspace*{-0.75\baselineskip}
	\label{tab:categories}
	\begin{tabular}{lp{12cm}}
		\toprule	
		\textbf{Main category} & \textbf{Subcategories} \\
		\midrule
		Active life & ATV Rentals/Tours, Airsoft, Amateur Sports Teams, Amusement Parks, Aquariums, Archery, Beaches, Bowling, Fitness \& Instruction, Golf, Hiking, Parks, Skydiving, Zoos, ... \\
		Arts and entertainment & Arcades, Art Galleries, Botanical Gardens, Casinos, Castles, Museums, Music Venues, LAN Centers, Performing Arts, Social Clubs, Street Art, Observatories, Wineries, Virtual Reality Centers, ... \\
		Restaurants and food & Coffee \& Tea, Desserts, Food Trucks, Internet Cafes, Specialty Food, Wineries, Afghan, African, American (Traditional), Asian Fusion, Beer Garden, Cafes, Comfort Food, Polish, ... \\
		Nightlife & Bar Crawl, Bars, Beer Gardens, Coffeeshops, Music Venues, Dance Clubs, Club Crawl, Pubs, Lounges, Hookah Bars, Gay Bars, Dance Restaurants, Pool Halls, Sports Bars, ... \\
		\bottomrule
	\end{tabular}
\end{table*}

\vspace*{-0.25\baselineskip}
\subsection{Collecting Information Needs}
\label{sec:infoneeds:collect}
\vspace*{-0.25\baselineskip}

Information needs were collected from two main sources: Yahoo! Answers (24 in total) and Reddit (88 in total).

In Yahoo! Answers, potential posts were identified by searching for recommendations specifically for each of the main categories (e.g., by issuing the query ``restaurant recommendations'').
A similar procedure was used for Reddit, but the search was restricted to specific subreddits that either correspond to a given location (e.g., ``Oslo'') or to a specific activity (e.g., ``hiking'').  A list of relevant subreddits was manually identified.\footnote{\url{https://github.com/iai-group/sigir2021-pointrec/blob/master/Reddit.md}} 

\vspace*{-0.25\baselineskip}
\subsection{Annotations and Metadata Extraction}
\label{sec:infoneeds:metadata}
\vspace*{-0.25\baselineskip}

\begin{table*}[t]  
  \centering
  \caption{Possible qualifiers expressing constraints for information needs.}
  \label{tbl:constraints} 
  \vspace*{-0.75\baselineskip}
  \begin{tabular}{lc ll} 
	\toprule
	\multirow{2}{*}{\textbf{Qualifier}} & \multirow{2}{*}{\textbf{Weight}} & \multicolumn{2}{c}{\textbf{Example}} \\
	\cline{3-4}
     & & \textbf{Constraints\_text} & \textbf{Constraints} \\  
    \midrule
	MUST (have/be) & 3 & "I want to see museums" & museums\\
    SHOULD (have/be) & 2 & "Good service is appreciated" & good service\\
    NICE\_TO (have/be) & 1 & "I like museums, monuments, and arts" & [museums, monuments, arts]\\
    \midrule
    NICE\_TO\_NOT (have/be) & -1 & "I don't prefer EDM, but I can live with it." & Electronic dance music\\
    SHOULD\_NOT (have/be) & -2 & "I'm not a fan of the big touristy spots" & tourist centric\\
    MUST\_NOT (have/be) & -3 & "I do not want to see museums" & museums\\
    \bottomrule
  \end{tabular}
\end{table*}

A unique property of the POINTREC collection is that information needs are manually annotated with rich metadata. In the beginning, ``raw'' information needs contain the following fields:
\begin{itemize}[leftmargin=2em]
	\item \textbf{URL}: Post where the information need originates from.
    \item \textbf{Title}: The title of the post.
    \item \textbf{Description}: The body of the post, describing the information need.
\end{itemize}
It is important to note that a single post may give rise to multiple information needs.  Specifically, it may happen that the user is seeking recommendations in multiple main categories, countries, or cities.  In such cases, a number of separate entries are created, which have identical URL, title, and description, but will differ in their metadata fields.
This information is also preserved in the unique \textbf{ID}, which follows a particular \texttt{AAAA-BBB-CC} format, where:
\begin{itemize}[leftmargin=2em]
	\item AAAA is four running digits to identify the post where the information need originates from;
	\item BBB is a counter that tracks if the originating post contains multiple information needs;
	\item CC is used to identify which category this information need belongs to (AL, AE, RF, or NL).
\end{itemize}
Next, information needs are enriched with the following metadata:
\begin{itemize}[leftmargin=2em]
    \item \textbf{Main Category}: This field specifies which main category the information need belongs to.  (By definition, each information need belongs to a single main category.)  Note that the main category is not necessarily explicitly listed, e.g., someone looking for hiking recommendations may not specify ``Active Life.'' The identification of main- and sub-categories is done manually, by reading the description and then matching it to one or several categories from Yelp.
    \item \textbf{Sub-categories}: An exhaustive list of all sub-categories that are relevant for the information need. 
    \item \textbf{Group}: A description of the group traveling.  For information needs where this is not specified, the field is left empty.  In all other cases, it is a contiguous piece of text extracted from the description (e.g., ``a group of 10'' or ``with my girlfriend'').
	\item \textbf{Constraints\_text}: A list of user preferences regarding the POI. This field is copied verbatim from the description (i.e., each element of the list is a contiguous piece of text in the description).  
	\item \textbf{Constraints}: Same as the field above, with the preferences structured into (qualifier, constraint) pairs.  Qualifiers come from a fixed set:  MUST, SHOULD, NICE\_TO, and their negative counterparts.  The natural language formulation of the constraint is possibly simplified.  For example, ``can't cost an arm and leg'' implies that they are not looking for anything too expensive, so it is interpreted as ``MUST\_NOT too expensive.'' Table~\ref{tbl:constraints} lists the possible qualifiers along with examples.
    \item \textbf{Positive examples}: Some information needs include POIs that the user likes, or, is interested in.  These POIs are listed here.
    \item \textbf{Negative examples}: Analogous to the Positive examples field, this is a listing of POIs that are explicitly disliked or indicated as not interesting.
    \item \textbf{Request\_text}: A specification of what the user is looking for, copied verbatim from the description.  It is worth to noting that this field may contain multiple contiguous pieces of text from the description.
    \item \textbf{Request}: Indicates the type of place/activity the user is looking for, e.g., ``restaurant,'' ``museum,'' ``hiking,'' or ``bars.''
    \item \textbf{Country}: The Alpha-2 code\footnote{\url{https://www.iso.org/glossary-for-iso-3166.html}} of the country the information need is located in.
	\item \textbf{State}: This field is primarily used for large countries (USA, Canada, and Australia).
    \item \textbf{City}: The name of the city targeted by the information need.\footnote{For some of the information needs, the city name and country name are not present in the title or description.  However, as the information needs were posted in forums dedicated to a specific city, this was inferred.}
\end{itemize}
Notice that there are two main types of metadata fields.
The first group is essentially a mark-up on the original description text, identifying certain types of preferences (group, constraints\_text, request\_text, positive/negative examples).
The second group of fields correspond to semantic annotations, which involve some degree of linguistic normalization (constraints, request) or restricting values to a controlled vocabulary (categories, country, state, city). 
Both types of metadata could potentially be extracted automatically; this, however, is left for future work.

Table~\ref{tab:example} shows an example information need, including the raw need (top) and the corresponding metadata enrichment (bottom).

\begin{table}[!t]
    \caption{Example information need (ID=0020-000-RF).}
    \label{tab:example}
	\vspace*{-0.75\baselineskip}
    \begin{tabular}{lp{6cm}}
    \toprule
    \multicolumn{2}{l}{\emph{\textbf{Raw information need}}} \\
    \midrule
    URL & https://answers.yahoo.com/question/index;... \\
	Title & Restaurant recommendation in London near Charing Cross? \\
    Description & I want to take my Dad out for a nice meal for a big birthday (really soon). Wondering if anyone can suggest anywhere? He is basically the typical grumpy old man, likes plenty on his plate, no waiters flapping over you (like good service!), no loud music, good ale on tap is a definite bonus and can't cost an arm and leg. It also has to be in walking distance of Charing Cross as he won't use the tube. Pubs or restaurants would be fine as long as they aren't crowded/noisy and meet the above criteria. Any suggestions? \\
    \midrule
    \multicolumn{2}{l}{\emph{\textbf{Metadata enrichment}}} \\
    \midrule
    Group & my father and I \\
    Main Category & Restaurants and Food \\
    Constraints\_text & He is basically the typical grumpy old man ... \\
    Constraints & MUST: {[}"Good service",
                "Walking distance of Charing Cross"{]} \\
	& SHOULD: {[}"Ale on tap"{]} \\
	& MUST\_NOT {[}"Loud music", "Crowded", "Noisy", "too expensive"{]} \\
	Request\_text & Restaurant recommendation in London near Charing Cross? \\
	Request & Restaurant \\
	Country & GB \\
	City & London \\
	\bottomrule
    \end{tabular}
	\vspace*{-1\baselineskip}
\end{table}

\subsection{Statistics}
\label{sec:infoneeds:stat}

A breakdown of the information needs across the main categories is shown in Table~\ref{tab:inpercat}.  It can be seen that information needs are roughly evenly distributed across the four main categories.

\begin{table}[t]
    \caption{Number of information needs per main category.}
    \label{tab:inpercat}
	\vspace*{-0.75\baselineskip}
    \begin{tabular}{lr}
    \toprule
    \textbf{Category} & \textbf{\#InfoNeeds} \\
    \midrule
         {[}AL{]} Active Life & 27 \\
         {[}AE{]} Arts \& Entertainment & 30 \\
         {[}RF{]} Restaurants and Food & 30 \\
         {[}NL{]} Nightlife & 25 \\
    \midrule
        Total & 112 \\
    \bottomrule
    \end{tabular}
\end{table}

\section{POI collection}

This section describes the process of creating our POI collection. Table~\ref{tab:example_poi} shows the information associated with an example POI in our collection.

\begin{table}[!t]
    \caption{Example POI (ID=11865).}
    \label{tab:example_poi}
	\vspace*{-0.75\baselineskip}
    \begin{tabular}{lp{5.5cm}}
    \toprule
    URL & https://www.yelp.com/biz... \\
    Image\_url & https://s3-media2.fl.yelpcdn.com/bphoto/... \\
	Name & L'Hu\^itrier \\
	Alias & l-hu\^itrier-paris-2 \\
	Main\_category & Restaurants and Food \\
	Sub\_categories & French, Seafood \\
	Rating & 4.0 \\
	Price & \euro{}\euro{}\euro{} \\
	Address & 16 rue Saussier Leroy \\
	City & Paris \\
	City\_from\_postal & Paris \\
	Zip\_code & 75017 \\
	Country\_code & FR \\
	State\_code & 75 \\
	Phone & +33140548344 \\
	Review\_count & 22 \\
	\midrule
    \end{tabular}
    \begin{tabular}{lp{6.6cm}}
	\emph{\textbf{Snippets}} \\
	\midrule
	~~~URL & https://www.tripadvisor.com/... \\
	~~~Title & L'Huitrier, Paris - Ternes - Menu, Prices \& Restaurant \\
	~~~Snippet & Order food online at L'Huitrier, Paris with Tripadvisor: See 320 unbiased reviews of L'Huitrier, ranked \#1,386 on Tripadvisor among 17,679 restaurants in Paris. \\
	\midrule
	~~~URL & https://www.viamichelin.com/... \\
	~~~Title & L'Hu\^itrier - Paris : a Michelin Guide restaurant \\
	~~~Snippet & L'Hu\^itrier - a restaurant from the Michelin Guide Find all the information you need on the L'Hu\^itrier restaurant: Michelin guide review, user reviews, cuisine, opening times, meal prices... L'Hu\^itrier - a Michelin Guide restaurant. \\
	\midrule
	... \\	
	\bottomrule
    \end{tabular}
	\vspace*{-0.75\baselineskip}
\end{table}

\subsection{Collecting POIs from Yelp}
\label{sec:poi:collect}

POIs were collected from Yelp, using their Fusion API (which is permitted for non-commercial uses\footnote{\url{https://www.Yelp.com/developers/api_terms}}). The POIs were primarily categorized within one of the four main categories mentioned in Section~\ref{sec:infoneeds:categories}, with the exception of a few manual recommendations, see~Section~\ref{sec:MTurk:Manual Recommendations}. The resulting collection consists of 695,053 POIs distributed across 23 countries.  A breakdown of POIs per country and per main category is provided in the online repository.\footnote{\url{https://github.com/iai-group/sigir2021-pointrec/blob/master/Statistics.md}}

\subsection{Linking to Geographical Knowledge Bases}
\label{sec:poi:geoKB_linking}

Two different geographical knowledge bases were used: Restcountries\footnote{\url{https://restcountries.eu}} and GeoNames.\footnote{\url{https://www.geonames.org}} Restcountries was used to retrieve the alpha2codes for every country in the world.  Then, using GeoNames, we identified the cities\footnote{Our definition of a city requires there to be at least 1,000 inhabitants in the given area. As a result, 138,088 cities were identified.} within the countries present in the set of information needs.

\subsection{Enrichment with Web Snippets}
\label{sec:poi:enrich}

We used Bing's Web Search API\footnote{\url{https://azure.microsoft.com/nb-no/services/cognitive-services/bing-web-search-api}} to further enrich a subset of the POI collection with short descriptions for each POI.  Specifically, this subset is comprised of all cities that are present in our set of information needs. 
As a query, we combined the POI name and city, to ensure that the web snippets corresponded to the correct POI.

\subsection{Data Cleaning and Post-processing}
\label{sec:poi:post-process}

The returned data from Yelp would occasionally include a very specific location as the city, e.g., a neighborhood within a city. To link these locations to their appropriate city name, we used the Zip\_code field together with GeoNames' postal code table. However, even the Zip\_code field of the returned data would sometimes be inconsistent.\footnote{In most cases however, at least one of these fields were correct.}

Additionally, some POIs were missing their main category. These were linked to their corresponding main category, by matching the sub-categories between the POI and the main categories.

\section{Recommendations and Relevance Assessments}
\label{sec:rec}

We follow the standard practice of pooling~\citep{Voorhees:2005:TEE} for collecting relevance annotations.  For each information need, a pool of candidate POIs is constructed by combining recommendations made by humans as well as by multiple configurations of a simple content-based recommender system.  These would be termed ``manual'' and ``automatic'' runs, respectively, following  TREC terminology.

\subsection{Manual Recommendations}
\label{sec:MTurk:Manual Recommendations}
Manual recommendations are solicited from two sources: extracted from forum posts and gathered via crowdsourcing.  

\subsubsection{Recommendations from Forum Posts}
\label{sec:rec:forum}

Given that information needs were collected from community QA sites, we can tap into the respective forum threads as a source of recommendations.  
Specifically, two authors of the paper manually processed the replies to the originating post of each information need to collect POI suggestions.

Note that some posts had no replies, thus no suggestions.  In fact, this was the case for 31 out of 112 information needs, with most of these originating from Yahoo! Answers.  On the contrary, some posts had too many replies.  Due to time restrictions, the five most relevant comments (following the default relevance-based ranking of the service) were used to gather POI suggestions.  The next step was to find the corresponding Yelp page, to collect the Yelp POI link, which was used later in a similar fashion to the links collected through crowdsourcing, see Section~\ref{sec:rec:mturk}.

\begin{figure}
    \centering
    \includegraphics[width=\columnwidth]{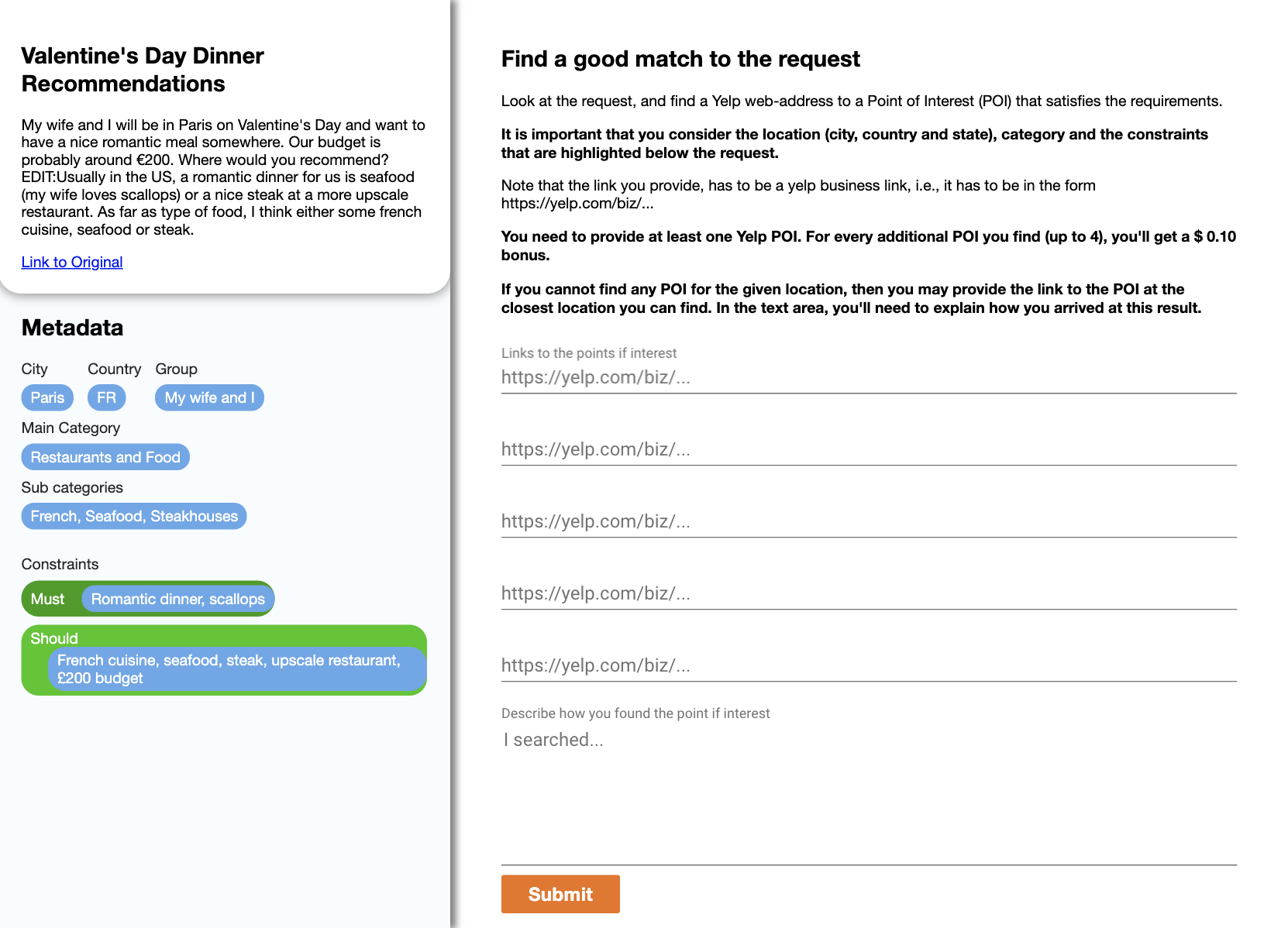}
    \caption{Crowdsourcing user interface for collecting POI recommendations.}
    \label{fig:mturk_task1}
	\vspace*{-0.75\baselineskip}
\end{figure}

\subsubsection{Recommendations Collected via Crowdsourcing}
\label{sec:rec:mturk}

Crowd workers, using the Amazon Mechanical Turk platform, were tasked with providing POI suggestions for a given information need.
Specifically, a user interface was developed (and refined in multiple iterations via pilot runs), where the most important metadata fields associated with the information need are highlighted; see Figure~\ref{fig:mturk_task1}.

Each worker was tasked to find at least 1 and at most 5 POIs.
The base payment was \$0.4, with a bonus of \$0.1 for each additional POI provided (on top of the mandatory one).
Since we aimed to collect as many suggestions as possible, each information need was processed by 5 different workers.
Workers were required to have more than 1,000 approved HITs, and a HIT approval rate above 95\%.
The submissions were manually checked by the paper authors.

\subsection{Automatic Recommendations}

Additionally, a simple content-based recommender system was implemented.  Specifically, the POI collection was indexed using Elasticsearch, using a separate field for each POI attribute.  
Then, three baseline rankers, incorporating various filtering criteria, were implemented by using structured queries. 
The main differences between these methods are summarized in Table~\ref{tab:baselines}; all three give a reward to POIs with a matching city.

\begin{itemize}[leftmargin=2em]
	\item \textbf{Baseline 1}: The main category of the POI was required to match the information need; categories have also been exploited in prior work to reduce the pool of candidates, e.g.,~\citep{Liu:2013:PPR,Sang:2015:ASC,Zhang:2015:GEG}. Additionally, POIs that match sub-categories will receive a higher relevance score. 
	\item \textbf{Baseline 2}: This method does not take into account main and sub-categories. Instead, it uses the request and constraints specified by the user and matches those against the text collected from web snippets. 
	\item \textbf{Baseline 3}: It is a combination of Baselines 1 and 2, and thus represents the strictest approach.
\end{itemize}

\begin{table}[t]
    \caption{Automatic recommender systems.}
	\vspace*{-0.75\baselineskip}
    \label{tab:baselines}
    \begin{tabular}{lcc}
    \toprule
    \textbf{System} & \textbf{Categories} & \textbf{Const. \& Reqs.} \\
    \midrule
        Baseline 1 & + & \\
        Baseline 2 &  & + \\
        Baseline 3 & + & + \\
    \bottomrule
    \end{tabular}
\end{table}

\subsection{Relevance Labeling}

\begin{figure}
    \centering
    \includegraphics[width=\columnwidth]{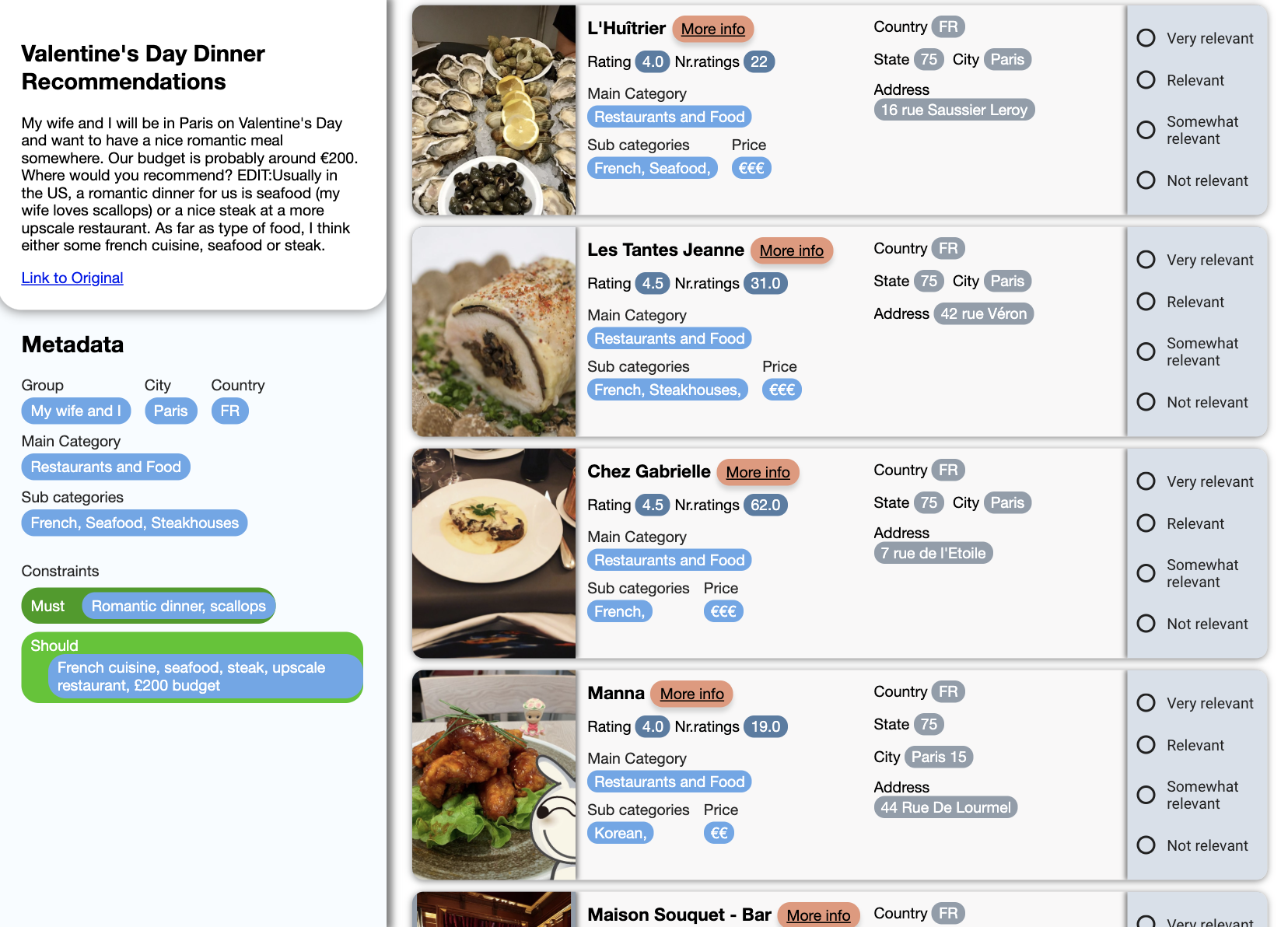}
    \vspace*{-1.5\baselineskip}
    \caption{Crowdsourcing UI for relevance assessments.}
    \label{fig:mturk_task2}
    \vspace*{-0.5\baselineskip}
\end{figure}

Relevance assessments were collected using crowdsourcing, using the interface shown in Figure~\ref{fig:mturk_task2}.
Each task consisted of 5 POI recommendations for a given information need and paid \$0.15. 
POI recommendations were pooled together from manual suggestions and the three automatic baselines (with up to top-50 results from each baseline).  The order of POIs was randomized to eliminate potential position bias.

Each information need-POI pair was judged by at least 3 workers on a 4-point relevance scale: 
\begin{itemize}
	\item A POI is \emph{highly relevant} (3), if it perfectly matches the request, with respect to location, category, and constraints.
	\item A POI is \emph{relevant} (2), if the location and category match, and the requested would likely be interested in it.
	\item A POI is \emph{somewhat relevant} (1), if the location matches, and the requester could possibly be interested in it.
	\item Otherwise, the POI is \emph{not relevant} (0).
\end{itemize}
In case of strong disagreement, that is, 2 points difference on the relevance scale, 2 additional judgments were collected.
Relevance labels were then consolidated by taking a majority vote.

Worker qualifications was the same as in Section~\ref{sec:rec:mturk}.
To ensure data quality and to avoid spammers, we used both gold answers (POI suggestions extracted manually from forums by the paper authors) and honeypots (injecting erroneous suggestions from a different main category). Workers with an error rate above 30\% were rejected.
The mean task completion time was 261sec.

\section{Baseline Results}

This section presents an experimental comparison of the baselines.
Our main measure is NDCG@5 (following the TREC Contextual Suggestions track~\citep{DeanHall:2015:OTC,Hashemi:2016:OTC}); we also report on NDCG@10. 
Additionally, we consider binary evaluation metrics, MAP and MRR, accepting only the highest relevance level (3) as relevant.

\begin{table}[t]
    \caption{Baseline results. Highest scores are boldfaced.}
    \label{tab:results}
	\vspace*{-0.75\baselineskip}
    \begin{tabular}{lcccc}
    \toprule
    \textbf{System} & \textbf{NDCG@5} & \textbf{NDCG@10} & \textbf{MRR} & \textbf{MAP} \\
    \midrule
        Baseline 1 & 0.6389 & 0.5812 & \textbf{0.5812} & \textbf{0.3304} \\
        Baseline 2 & 0.4109 & 0.3979 & 0.2814 & 0.0667 \\
        Baseline 3 & \textbf{0.6784} & \textbf{0.6573} & 0.5535 & 0.2506 \\
    \bottomrule
    \end{tabular}
\end{table}

\begin{figure}
    \centering
    \includegraphics[width=0.8\columnwidth]{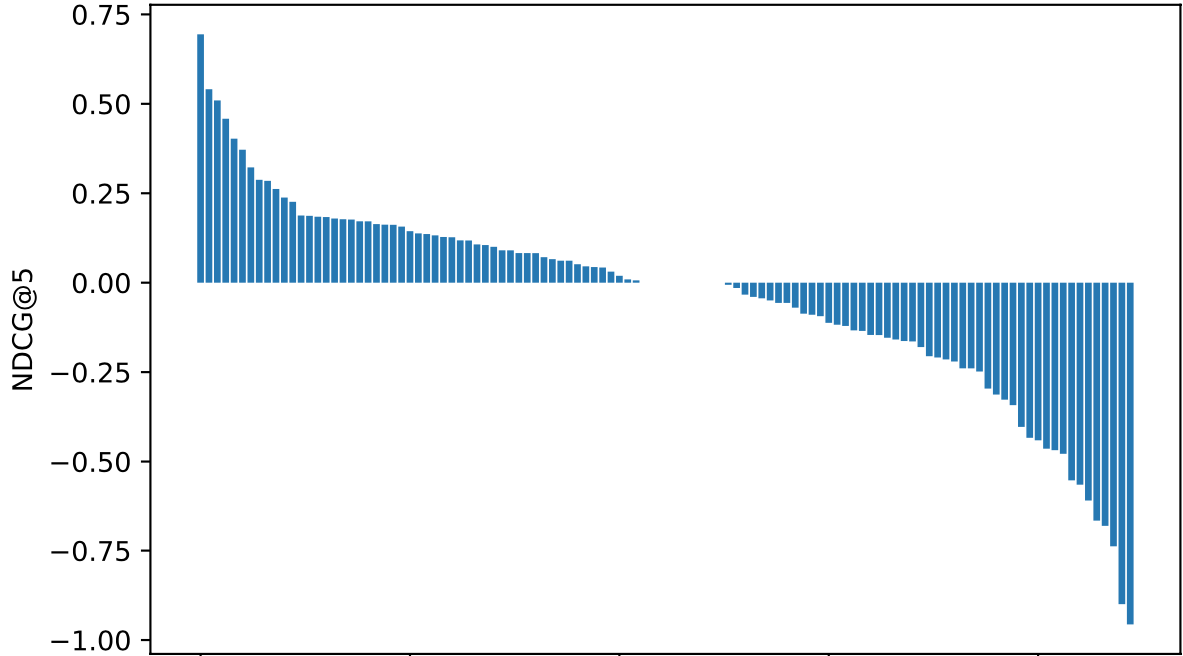}
    \caption{NDCG@5 difference between Baseline 1 and Baseline 3. Each bar corresponds to an information need. Positive values indicate that Baseline 1 performs better, negative values indicate that Baseline 3 is more effective.}
    \label{fig:topiclevel}
    \vspace*{-0.5\baselineskip}
\end{figure}

Table~\ref{tab:results} displays the results.
We observe that the strictest method (Baseline 3) performs best in terms of graded relevance measures, while Baseline 1 is the most effective in terms of binary metrics.
Figure~\ref{fig:topiclevel} shows a comparison between Baselines 1 and 3 on the level of individual information needs.
We find that on about half of the information needs Baseline 1 works better, while on the other half Baseline 3 is preferred.
Recall that both methods consider categories, while Baseline 3 additionally matches the requests and constraints as well against web snippets.
These results show that constraints can potentially aid retrieval, but more sophisticated approaches are needed in order to effectively incorporate them.

\section{Conclusion and future work}

This paper has introduced the POINTREC test collection for contextual POI recommendation.  It consists of a collection of POIs across 23 countries, a set of 112 information needs originating from real users, and graded relevance assessments collected via crowdsourcing.  Additionally, we have developed and experimentally compared three variants of a simple retrieval-based baseline approach.

This collection is primarily aimed at facilitating research on POI recommendation in a narrative-based scenario, where preferences and constraints are given in natural language.  These have been manually identified and provided as metadata annotations, and have the potential to improve the quality of recommendations.
The automatic extraction, representation, and interpretation of user preferences is an interesting and challenging natural language understanding task on its own account that is not exclusive to our setting; it is also a central component, for instance, in the context of conversational assistants (see, e.g.,~\citep{Balog:2021:OIM}).
Our annotated set of information needs provides a resource for research in this area.

\balance

\bibliographystyle{ACM-Reference-Format}
\bibliography{sigir2021-poi}


\end{document}